\documentclass[aps,showpacs,preprintnumbers,amsmath, amssymb]{revtex4}

\usepackage{float}
\usepackage{graphics,epsfig}
\usepackage{graphicx}
\usepackage{dcolumn}
\usepackage{bm}

\begin{document}
\baselineskip=0.8 cm
\title{{\bf On analytical study of holographic superconductors with Born-Infeld electrodynamics}}

\author{Chuyu Lai$^{1,2}$, Qiyuan Pan$^{1,2,3}$\footnote{panqiyuan@126.com}, Jiliang Jing$^{1,2}$\footnote{jljing@hunnu.edu.cn}
and Yongjiu Wang$^{1,2}$\footnote{wyj@hunnu.edu.cn}}
\affiliation{$^{1}$Institute of Physics and Department of Physics,
Hunan Normal University, Changsha, Hunan 410081, China}
\affiliation{$^{2}$ Key Laboratory of Low Dimensional Quantum
Structures and Quantum Control of Ministry of Education, Hunan
Normal University, Changsha, Hunan 410081, China}
\affiliation{$^{3}$ Instituto de F\'{\i}sica, Universidade de
S\~{a}o Paulo, CP 66318, S\~{a}o Paulo 05315-970, Brazil}

\vspace*{0.2cm}
\begin{abstract}
\baselineskip=0.6 cm
\begin{center}
{\bf Abstract}
\end{center}

Based on the Sturm-Liouville eigenvalue problem, Banerjee \emph{et
al.} proposed a perturbative approach to analytically investigate
the properties of the ($2+1$)-dimensional superconductor with
Born-Infeld electrodynamics [Phys. Rev. D {\bf 87}, 104001 (2013)].
By introducing an iterative procedure, we will further improve the
analytical results and the consistency with the numerical findings,
and can easily extend the analytical study to the higher-dimensional
superconductor with Born-Infeld electrodynamics. We observe that the
higher Born-Infeld corrections make it harder for the condensation
to form but do not affect the critical phenomena of the system. Our
analytical results can be used to back up the numerical computations
for the holographic superconductors with various condensates in
Born-Infeld electrodynamics.

\end{abstract}

%\keywords{AdS/CFT correspondence, Holographic superconductors, Born-Infeld electrodynamics}

\pacs{11.25.Tq, 04.70.Bw, 74.20.-z}\maketitle
\newpage
\vspace*{0.2cm}

\section{Introduction}

As one of the most significant developments in fundamental physics
in the last one decade, the anti-de Sitter/conformal field theories
(AdS/CFT) correspondence \cite{Maldacena,Witten,Gubser1998} allows
to describe the strongly coupled conformal field theories through a
weakly coupled dual gravitational description. A recent interesting
application of such a holography is constructing of a model of a
high $T_{c}$ superconductor, for reviews, see Refs.
\cite{HartnollRev,HerzogRev,HorowitzRev,CaiRev} and references
therein. It was found that the instability of the bulk black hole
corresponds to a second order phase transition from normal state to
superconducting state which brings the spontaneous U(1) symmetry
breaking \cite{GubserPRD78}, and the properties of a
($2+1$)-dimensional superconductor can indeed be reproduced in the
($3+1$)-dimensional holographic dual model based on the framework of
usual Maxwell electrodynamics \cite{HartnollPRL101}. In order to
understand the influences of the $1/N$ or $1/\lambda$ ($\lambda$ is
the 't Hooft coupling) corrections on the holographic dual models,
it is of great interest to consider the holographic superconductor
models with the nonlinear electrodynamics since the nonlinear
electrodynamics essentially implies the higher derivative
corrections of the gauge field \cite{HendiJHEP}. Jing and Chen
introduced the first holographic superconductor model in Born-Infeld
electrodynamics and observed that the nonlinear Born-Infeld
corrections will make it harder for the scalar condensation to form
\cite{JS2010}. Along this line, there have been accumulated interest
to study various holographic dual models with the nonlinear
electrodynamics
\cite{JLQS2012,JingJHEP,PJWPRD,SDSL2012JHEP156,LeeEPJC,LPW2012,
Roychowdhury,BGQX,JPCPLB,ZPCJNPB,YaoJing,DLAP,SGMPLA}.

In most cases, the holographic dual models were studied numerically.
In order to back up numerical results and gain more insights into
the properties of the holographic superconductors, Siopsis \emph{et
al.} developed the variational method for the Sturm-Liouville (S-L)
eigenvalue problem to analytically calculate the critical exponent
near the critical temperature and found that the analytical results
obtained by this way are in good agreement with the numerical
findings \cite{Siopsis,SiopsisBF}. Generalized to study the
holographic insulator/superconductor phase transition
\cite{Cai-Li-Zhang}, this method can clearly present the
condensation and critical phenomena of the system at the critical
point in AdS soliton background.

More recently, Gangopadhyay and Roychowdhury extended the S-L method
to investigate the properties of the ($2+1$)-dimensional
superconductor with Born-Infeld electrodynamics by introducing a
perturbative technique, and observed that the analytical results
agree well with the existing numerical results for the condensation
operator $\langle\mathcal{O}_{-}\rangle$ \cite{SDSL2012}. For the
operator $\langle\mathcal{O}_{+}\rangle$, Banerjee \emph{et al.}
improved the perturbative approach and explored the effect of the
Born-Infeld electrodynamics on the ($2+1$)-dimensional
superconductor \cite{BGRLPRD2013}. However, comparing with the case
of $\langle\mathcal{O}_{-}\rangle$ \cite{SDSL2012}, we find that for
the operator $\langle\mathcal{O}_{+}\rangle$ the agreement of the
analytical result with the numerical calculation is not so good, for
example in the case of the Born-Infeld parameter $b=0.3$
\cite{BGRLPRD2013}, the difference between the analytical and
numerical values is $22.1\%$! Furthermore, this perturbative
approach is not very valid to study the higher-dimensional
superconductor with Born-Infeld electrodynamics. Thus, the
motivation for completing this work is two fold. On one level, it is
worthwhile to reduce the disparity between the analytical and
numerical results for the operator $\langle\mathcal{O}_{+}\rangle$,
and further improve the analytical results and the consistency with
the numerical findings. On another more speculative level, it would
be important to develop a more general analytical technique which
can be used to study systematically the $d$-dimensional
superconductors with Born-Infeld electrodynamics and see some
general features for the effects of the higher derivative
corrections to the gauge field on the holographic dual models. In
order to avoid the complex computation, in this work we will
concentrate on the probe limit where the backreaction of matter
fields on the spacetime metric is neglected.

The plan of the work is the following. In Sec. II we will introduce
the holographic superconductor models with Born-Infeld
electrodynamics in the $(d+1)$-dimensional AdS black hole
background. In Sec. III we will improve the perturbative approach
proposed in \cite{BGRLPRD2013} and give an analytical investigation
of the holographic superconductors with Born-Infeld electrodynamics
by using the S-L method. We will conclude in the last section with
our main results.

\section{Holographic superconductors with Born-Infeld electrodynamics}

We begin with the background of the $(d+1)$-dimensional planar
Schwarzschild-AdS black hole
\begin{equation}
ds^2=-r^{2}f(r)dt^2+\frac{dr^2}{r^2f(r)}+r^2\sum_{i=1}^{d-1}dx_i^2,
\end{equation}
where $f(r)=1-r_+^{d}/r^{d}$ with the radius of the event horizon $r_{+}$. For convenience, we have set the AdS
radius $L=1$. The Hawking temperature of the black hole is determined by
\begin{eqnarray}
T=\frac{dr_{+}}{4\pi},
\end{eqnarray}
which will be interpreted as the temperature of the CFT.

Working in the probe limit, we consider the Born-Infeld electrodynamics and the charged complex scalar field coupled via the action
\begin{eqnarray}\label{System}
S=\int
d^{d+1}x\sqrt{-g}\left[\frac{1}{b}\left(1-\sqrt{1+\frac{1}{2}bF^{2}}\right)
-|\nabla\psi-iA\psi|^{2}-m^2|\psi|^2\right],
\end{eqnarray}
with the quadratic term $F^{2}=F_{\mu\nu}F^{\mu\nu}$. When the
Born-Infeld parameter $b\rightarrow0$, the model (\ref{System})
reduces to the standard holographic superconductors investigated in
\cite{HartnollPRL101,HorowitzPRD78}.

With the ansatz of the matter fields as $\psi=|\psi|$, $A_{t}=\phi$
where $\psi$ and $\phi$ are both real functions of $r$ only, we can
arrive at the following equations of motion for the scalar field
$\psi$ and the gauge field $\phi$
\begin{eqnarray}
&&\psi^{\prime\prime}+\left(
\frac{1+d}{r}+\frac{f^\prime}{f}\right)\psi^\prime
+\left(\frac{\phi^2}{r^4f^2}-\frac{m^2}{r^2f}\right)\psi=0,
\label{BHPsir}
\end{eqnarray}
\begin{eqnarray}
\phi^{\prime\prime}+\frac{d-1}{r}\left(1-b\phi^{\prime
2}\right)\phi^\prime-\frac{2\psi^{2}}{r^2f}\left(1-b\phi^{\prime
2}\right)^{3/2}\phi=0,\label{BHPhir}
\end{eqnarray}
where the prime denotes the derivative with respect to $r$.

Applying the S-L method to analytically study the properties of the
holographic superconductors with Born-Infeld electrodynamics, we
will introduce a new variable $z=r_{+}/r$ and rewrite the equations
of motion (\ref{BHPsir}) and  (\ref{BHPhir}) into
\begin{eqnarray}
&&\psi^{\prime\prime}+\left(
\frac{1-d}{z}+\frac{f^\prime}{f}\right)\psi^\prime
+\left(\frac{\phi^2}{r_{+}^2f^2}-\frac{m^2}{z^2f}\right)\psi=0,
\label{BHPsiz}
\end{eqnarray}
\begin{eqnarray}
\phi^{\prime\prime}+\frac{1}{z}\left[(3-d)+\frac{b(d-1)z^{4}}{r_{+}^{2}}\phi^{\prime
2}\right]\phi^\prime-\frac{2\psi^{2}}{z^2f}\left(1-\frac{bz^{4}}{r_{+}^{2}}\phi^{\prime
2}\right)^{3/2}\phi=0,\label{BHPhiz}
\end{eqnarray}
with $f=1-z^{d}$. Here and hereafter the prime denotes the
derivative with respect to $z$.

In order to get the solutions in the superconducting phase, we have
to impose the appropriate boundary conditions for $\psi$ and $\phi$.
At the event horizon $z=1$ of the black hole, the regularity gives
the boundary conditions
\begin{eqnarray}
\psi(1)=-\frac{d}{m^{2}}\psi^\prime(1)\,,\hspace{0.5cm} \phi(1)=0\,.
\label{horizon}
\end{eqnarray}
Near the AdS boundary $z\rightarrow0$, the asymptotic behaviors of
the solutions are
\begin{eqnarray}
\psi=\frac{\psi_{-}}{r_{+}^{\Delta_{-}}}z^{\Delta_{-}}+\frac{\psi_{+}}{r_{+}^{\Delta_{+}}}z^{\Delta_{+}}\,,\hspace{0.5cm}
\phi=\mu-\frac{\rho}{r_{+}^{d-2}}z^{d-2}\,, \label{infinity}
\end{eqnarray}
where $\Delta_\pm=(d\pm\sqrt{d^{2}+4m^{2}})/2$ is the conformal
dimension of the scalar operator dual to the bulk scalar field,
$\mu$ and $\rho$ are interpreted as the chemical potential and
charge density in the dual field theory respectively. It should be
pointed out that, provided $\Delta_{-}$ is larger than the unitarity
bound, both $\psi_{-}$ and $\psi_{+}$ can be normalizable and they
can be used to define operators in the dual field theory according
to the AdS/CFT correspondence,
$\psi_{-}=\langle\mathcal{O}_{-}\rangle$ and
$\psi_{+}=\langle\mathcal{O}_{+}\rangle$, respectively. Just as in
Refs. \cite{HartnollPRL101,HorowitzPRD78}, we will impose boundary
condition that either $\psi_{-}$ or $\psi_{+}$ vanishes. In this
work, we impose boundary condition $\psi_{-}=0$ since we concentrate
on the condensate for the operator $\langle\mathcal{O}_{+}\rangle$.
For clarity, we set
$\langle\mathcal{O}\rangle=\langle\mathcal{O}_{+}\rangle$ and $
\Delta=\Delta_{+}$ in the following discussion.

\section{Analytical study of holographic superconductors with Born-Infeld electrodynamics}

Here we will improve the perturbative approach proposed in
\cite{BGRLPRD2013} and use the S-L method \cite{Siopsis} to
analytically discuss the properties of the $d$-dimensional
superconductor phase transition with Born-Infeld electrodynamics. We
will investigate the relation between critical temperature and
charge density as well as the critical exponent of condensation
operators, and examine the effect of the Born-Infeld parameter.

\subsection{Critical temperature}

At the critical temperature $T_{c}$, the scalar field $\psi=0$.
Thus, near the critical point the equation of motion (\ref{BHPhiz})
for the gauge field $\phi$ becomes
\begin{eqnarray}
\phi^{\prime\prime}+\frac{1}{z}\left[(3-d)+\frac{b(d-1)z^{4}}{r_{+c}^{2}}\phi^{\prime
2}\right]\phi^\prime=0,\label{NESWPhiCritical}
\end{eqnarray}
where $r_{+c}$ is the radius of the horizon at the critical point.
Defining $\xi(z)=\phi'(z)$, we can obtain
\begin{eqnarray}
\xi^{\prime}+\frac{3-d}{z}\xi=\frac{b(1-d)z^{3}}{r_{+c}^{2}}\xi^{3},\label{XiCritical}
\end{eqnarray}
which is the special case of Bernoulli's Equation
$y'(x)+f(x)y=g(x)y^{n}$ \cite{Chow} for $n=3$. Considering that the
boundary condition (\ref{infinity}) for $\phi$, we can get the
solution to Eq. (\ref{XiCritical})
\begin{eqnarray}
\xi(z)=\phi'(z)=-\frac{\lambda
r_{+c}(d-2)z^{d-3}}{\sqrt{1+(d-2)^{2}b\lambda^{2}z^{2(d-1)}}},
\label{XiSolution}
\end{eqnarray}
which leads to the expression
\begin{eqnarray}
\phi(z)=\lambda r_{+c}\zeta(z), \label{PhiSolution}
\end{eqnarray}
with
\begin{eqnarray}
\zeta(z)=\int^{1}_{z}\frac{(d-2)\tilde{z}^{d-3}}{\sqrt{1+(d-2)^{2}b\lambda^{2}\tilde{z}^{2(d-1)}}}d\tilde{z},
\label{ZetaSolution}
\end{eqnarray}
where we have set $\lambda=\rho/r^{d-1}_{+c}$ and used the fact that
$\phi(1)=0$.

Obviously, the integral in (\ref{ZetaSolution}) is not doable
exactly. Just as in Refs. \cite{SDSL2012,BGRLPRD2013}, we will
perform a perturbative expansion of $(d-2)^{2}b\lambda^{2}$. In
order to simplify the following calculation, we will express the
Born-Infeld parameter $b$ as
\begin{eqnarray}\label{BIbn}
b_{n}=n\Delta b,~~~n=0,1,2,\cdot\cdot\cdot,
\end{eqnarray}
where $\Delta b=b_{n+1}-b_{n}$ is the step size of our iterative
procedure. Considering the fact that
\begin{eqnarray}\label{bLambda}
(d-2)^{2}b\lambda^{2}=(d-2)^{2}b_{n}\lambda^{2}=(d-2)^{2}b_{n}(\lambda^{2}|_{b_{n-1}})+0[(\Delta
b)^{2}],
\end{eqnarray}
where we have set $b_{-1}=0$ and $\lambda^{2}|_{b_{-1}}=0$, we will
discuss the following two cases (note that the variable $z$ has a
range $0\leq z\leq1$):

\textit{Case 1.} If $(d-2)^{2}b_{n}(\lambda^{2}|_{b_{n-1}})<1$, we
have
\begin{eqnarray}
&\zeta(z)&=\zeta_{1}(z)\approx\int^{1}_{z}(d-2)\tilde{z}^{d-3}\left[1-\frac{(d-2)^{2}b_{n}(\lambda^{2}|_{b_{n-1}})\tilde{z}^{2(d-1)}}{2}\right]d\tilde{z}
\nonumber\\
&&=(1-z^{d-2})+\frac{(d-2)^{3}b_{n}(\lambda^{2}|_{b_{n-1}})}{2(4-3d)}(1-z^{3d-4}).
\label{ZetaCase1}
\end{eqnarray}

\textit{Case 2.} If $(d-2)^{2}b_{n}(\lambda^{2}|_{b_{n-1}})>1$, we
set $(d-2)^{2}b_{n}(\lambda^{2}|_{b_{n-1}})\Lambda^{2(d-1)}=1$ for
$z=\Lambda$. Obviously, we find that
$(d-2)^{2}b_{n}(\lambda^{2}|_{b_{n-1}})z^{2(d-1)}<1$ for
$z<\Lambda<1$, which results in
\begin{eqnarray}
&\zeta(z)&=\zeta_{2A}(z)\approx\int^{\Lambda}_{z}(d-2)\tilde{z}^{d-3}\left[1-\frac{(d-2)^{2}b_{n}(\lambda^{2}|_{b_{n-1}})\tilde{z}^{2(d-1)}}{2}\right]d\tilde{z}
\nonumber\\
&&\qquad\qquad+\int^{1}_{\Lambda}\frac{1}{\sqrt{b_{n}}(\lambda|_{b_{n-1}})\tilde{z}^{2}}\left[1-
\frac{1}{2(d-2)^{2}b_{n}(\lambda^{2}|_{b_{n-1}})\tilde{z}^{2(d-1)}}\right]d\tilde{z}
\nonumber\\
&&=-z^{d-2}+\frac{(d-2)z^{3d-4}}{2(3d-4)\Lambda^{2(d-1)}}+\frac{3(d-1)[6+d(4d-9)]}{2(2d-1)(3d-4)}\Lambda^{d-2}
+(d-2)\Lambda^{d-1}\left[\frac{\Lambda^{2(d-1)}}{2(2d-1)}-1\right],
\label{ZetaCase2A}
\end{eqnarray}
and $(d-2)^{2}b_{n}(\lambda^{2}|_{b_{n-1}})z^{2(d-1)}>1$ for
$\Lambda< z\leq1$, which leads to
\begin{eqnarray}
&\zeta(z)&=\zeta_{2B}(z)\approx\int^{1}_{z}\frac{1}{\sqrt{b_{n}}(\lambda|_{b_{n-1}})\tilde{z}^{2}}\left[1-
\frac{1}{2(d-2)^{2}b_{n}(\lambda^{2}|_{b_{n-1}})\tilde{z}^{2(d-1)}}\right]d\tilde{z}
\nonumber\\
&&=(d-2)\Lambda^{d-1}\left[\frac{\Lambda^{2(d-1)}}{2(2d-1)}\left(1-z^{1-2d}\right)+\frac{1}{z}-1\right].
\label{ZetaCase2B}
\end{eqnarray}
It should be noted that in both cases we observe that $\zeta(1)=0$
from (\ref{ZetaCase1}) and (\ref{ZetaCase2B}), which is consistent
with the boundary condition $\phi(1)=0$ given in (\ref{horizon}).

Introducing a trial function $F(z)$ near the boundary $z=0$ as
\begin{eqnarray}
\psi(z)\sim\frac{\langle\mathcal{O}\rangle}{r_{+}^{\Delta}}
z^{\Delta}F(z), \label{BintroduceF}
\end{eqnarray}
with the boundary conditions $F(0)=1$ and $F'(0)=0$, from Eq.
(\ref{BHPsiz}) we can obtain the equation of motion for $F(z)$
\begin{eqnarray}\label{BFEoM}
(TF^{\prime})^{\prime}+T\left(P+\lambda^2Q\zeta^{2}\right)F=0,
\end{eqnarray}
with
\begin{eqnarray}
T=z^{1+2\Delta-d}(1-z^{d}),~~P=\frac{\Delta(\Delta-d)}{z^{2}}+\frac{\Delta
f'}{z f}-\frac{m^{2}}{z^{2}f},~~ Q=\frac{1}{f^{2}}.
\end{eqnarray}
According to the S-L eigenvalue problem \cite{Gelfand-Fomin}, we
deduce the eigenvalue $\lambda$ minimizes the expression
\begin{eqnarray}\label{lambdaeigenvalueCase1}
\lambda^{2}=\frac{\int^{1}_{0}T\left(F'^{2}-PF^{2}\right)dz}{\int^{1}_{0}TQ\zeta_{1}^{2}F^2dz}\
, &  \quad {\rm for} \ (d-2)^{2}b_{n}(\lambda^{2}|_{b_{n-1}})<1,
\end{eqnarray}
and
\begin{eqnarray}\label{lambdaeigenvalueCase2}
\lambda^{2}=\frac{\int^{1}_{0}T\left(F'^{2}-PF^{2}\right)dz}
{\int^{\Lambda}_{0}TQ\zeta_{2A}^{2}F^2dz+\int^{1}_{\Lambda}TQ\zeta_{2B}^{2}F^2dz}\
, &  \quad {\rm for} \ (d-2)^{2}b_{n}(\lambda^{2}|_{b_{n-1}})>1.
\end{eqnarray}
Using Eqs. (\ref{lambdaeigenvalueCase1}) and
(\ref{lambdaeigenvalueCase2}) to compute the minimum eigenvalue of
$\lambda^{2}$, we can obtain the critical temperature $T_{c}$ for
different Born-Infeld parameter $b$, spacetime dimension $d$ and
mass of the scalar field $m$ from the following relation
\begin{eqnarray}\label{CTTc}
T_{c}=\frac{d}{4\pi}\left(\frac{\rho}{\lambda_{min}}\right)^{\frac{1}{d-1}}.
\end{eqnarray}
In the following calculation, we will assume the trial function to
be $F(z)=1-az^{2}$ with a constant $a$.

As an example, we will study the case for $d=3$ and $m^{2}L^2=-2$
with the chosen values of the Born-Infeld parameter $b$. Setting
$\Delta b=0.1$, for $b_{0}=0$ we use Eq.
(\ref{lambdaeigenvalueCase1}) and get
\begin{eqnarray}
\lambda^{2}=\frac{4(15-20a+12a^{2})}{10(9-\sqrt{3}\pi-3\ln3)+10(13-12\ln3)a+(10\sqrt{3}\pi-21-30\ln3)a^{2}},
\end{eqnarray}
whose minimum is $\lambda^{2}|_{b_{0}}=17.31$ at $a=0.6016$.
According to Eq. (\ref{CTTc}), we can easily obtain the critical
temperature $T_{c}=0.1170\rho^{1/2}$, which is in good agreement
with the numerical result $T_{c}=0.1184\rho^{1/2}$
\cite{HartnollPRL101}. For $b_{1}=0.1$, we can easily have
$b_{1}(\lambda^{2}|_{b_{0}})>1$ and
$\Lambda=[b_{1}(\lambda^{2}|_{b_{0}})]^{-1/4}=0.8718$. Using Eq.
(\ref{lambdaeigenvalueCase2}) we arrive at
\begin{eqnarray}
\lambda^{2}=\frac{1-\frac{4a}{3}+\frac{4a^{2}}{5}}{0.02060-0.01199a+0.002659a^{2}},
\end{eqnarray}
whose minimum is $\lambda^{2}|_{b_{1}}=33.84$ at $a=0.6532$. So the
critical temperature $T_{c}=0.09898\rho^{1/2}$, which also agrees
well with the numerical finding $T_{c}=0.1007\rho^{1/2}$
\cite{JS2010}. For $b_{1}=0.2$, we still have
$b_{2}(\lambda^{2}|_{b_{1}})>1$ and
$\Lambda=[b_{2}(\lambda^{2}|_{b_{1}})]^{-1/4}=0.6200$. With the help
of Eq. (\ref{lambdaeigenvalueCase2}) we obtain
\begin{eqnarray}
\lambda^{2}=\frac{1-\frac{4a}{3}+\frac{4a^{2}}{5}}{0.01176-0.006582a+0.001450a^{2}},
\end{eqnarray}
whose minimum is $\lambda^{2}|_{b_{2}}=58.19$ at $a=0.6640$.
Therefore the critical temperature $T_{c}=0.08644\rho^{1/2}$, which
is again consistent with the numerical result
$T_{c}=0.08566\rho^{1/2}$ \cite{JS2010}. For other values of $b$,
the similar iterative procedure can be applied to give the
analytical result for the critical temperature.

\begin{table}[ht]
\begin{center}
\caption{\label{BICriticalTcD3} The critical temperature $T_{c}$
obtained by the analytical S-L method and from numerical calculation
\cite{JS2010} for the chosen values of the Born-Infeld parameter $b$
in the case of 4-dimensional AdS black hole background. Here we fix
the mass of the scalar field by $m^{2}L^2=-2$ and the step size by
$\Delta b=0.1$.}
\begin{tabular}{c c c c c c c}
         \hline \hline
$b$ & 0 & 0.1 & 0.2 & 0.3
        \\
        \hline
~~~~$Analytical$~~~~~~~~&~~~~~~~~$0.1170\rho^{1/2}$~~~~~~~~&~~~~~~~~$0.09898\rho^{1/2}$
~~~~~~~~&~~~~~~~~$0.08644\rho^{1/2}$~~~~~~~~&~~~~~~~~$0.07586\rho^{1/2}$~~~~
          \\
~~~~$Numerical$~~~~~~~~&~~~~~~~~$0.1184\rho^{1/2}$~~~~~~~~&~~~~~~~~$0.1007\rho^{1/2}$
~~~~~~~~&~~~~~~~~$0.08566\rho^{1/2}$~~~~~~~~&~~~~~~~~$0.07292\rho^{1/2}$~~~~
          \\
        \hline \hline
\end{tabular}
\end{center}
\end{table}

In Table \ref{BICriticalTcD3}, we provide the critical temperature
$T_{c}$ of the chosen parameter $b$ with the scalar operator
$\langle\mathcal{O}\rangle=\langle\mathcal{O}_{+}\rangle$ for the
($2+1$)-dimensional superconductor if we fix the mass of the scalar
field by $m^{2}L^2=-2$ and the step size by $\Delta b=0.1$. From
Table \ref{BICriticalTcD3}, we observe that the differences between
the analytical and numerical values are within $4.1\%$. Compared
with the analytical results given in Table 1 of Ref.
\cite{BGRLPRD2013}, the iterative procedure can further improve our
analytical results and improve the consistency with the numerical
findings.

Extending the investigation to the ($3+1$)-dimensional
superconductor, in Table \ref{BICriticalTcD4} we also give the
critical temperature $T_{c}$ for the scalar operator
$\langle\mathcal{O}\rangle=\langle\mathcal{O}_{+}\rangle$ when we
fix the mass of the scalar field $m^{2}L^2=-3$ for different
Born-Infeld parameter $b$ by choosing the step size $\Delta b=0.05$
and $0.025$, respectively. Obviously, for the case of $\Delta
b=0.025$ the agreement of the analytical results derived from S-L
method with the numerical calculation is impressive. Thus, we argue
that, even in the higher dimension, the analytical results derived
from the S-L method are in very good agreement with the numerical
calculation. Furthermore, reducing the step size $\Delta b$
reasonably, we can improve the analytical result and get the
critical temperature more consistent with the numerical result.

\begin{table}[ht]
\begin{center}
\caption{\label{BICriticalTcD4} The critical temperature $T_{c}$
with the chosen values of the Born-Infeld parameter $b$ and the step
size $\Delta b$ in the case of 5-dimensional AdS black hole
background. Here we fix the mass of the scalar field by
$m^{2}L^2=-3$.}
\begin{tabular}{c c c c c c c}
         \hline \hline
$b$ & 0 & 0.1 & 0.2 & 0.3
        \\
        \hline
~~~~$Analytical(\Delta
b=0.05)$~~~~~~&~~~~~~$0.1962\rho^{1/3}$~~~~~~&~~~~~~$0.1460\rho^{1/3}$
~~~~~~&~~~~~~$0.1091\rho^{1/3}$~~~~~~&~~~~~~$0.07866\rho^{1/3}$~~~~
          \\
~~~~$Analytical(\Delta
b=0.025)$~~~~~~&~~~~~~$0.1962\rho^{1/3}$~~~~~~&~~~~~~$0.1329\rho^{1/3}$
~~~~~~&~~~~~~$0.08754\rho^{1/3}$~~~~~~&~~~~~~$0.05195\rho^{1/3}$~~~~
          \\
~~~~$Numerical$~~~~~~&~~~~~~$0.1980\rho^{1/3}$~~~~~~&~~~~~~$0.1275\rho^{1/3}$
~~~~~~&~~~~~~$0.08298\rho^{1/3}$~~~~~~&~~~~~~$0.05292\rho^{1/3}$~~~~
          \\
        \hline \hline
\end{tabular}
\end{center}
\end{table}

From Tables \ref{BICriticalTcD3} and \ref{BICriticalTcD4}, we point
out that the critical temperature $T_{c}$ decreases as the
Born-Infeld parameter $b$ increases for the fixed scalar field mass
and spacetime dimension, which supports the numerical computation
found in Refs. \cite{JS2010,JLQS2012,ZPCJNPB}. It is shown that the
higher Born-Infeld electrodynamics corrections will make the scalar
hair more difficult to be developed. On the other hand, the
consistency between the analytical and numerical results indicates
that the S-L method is a powerful analytical way to investigate the
holographic superconductor with various condensates even when we
take the Born-Infeld electrodynamics into account.

\subsection{Critical phenomena}
Since the condensation for the scalar operator
$\langle\mathcal{O}\rangle$ is so small when $T \rightarrow T_c$, we
can expand $\phi(z)$ in $\langle\mathcal{O}\rangle$ near the
boundary $z=0$ as
\begin{eqnarray}\label{PhiExpandNearTc}
\frac{\phi(z)}{r_+}=\lambda\zeta(z)+\frac{\langle\mathcal{O}\rangle^2}{r_+^{2\Delta}}\chi(z)+\cdot\cdot\cdot,
\end{eqnarray}
with the boundary conditions $\chi(1)=0$ and $\chi'(1)=0$
\cite{Siopsis,ZPJ2015,Li-Cai-Zhang}. Thus, substituting the
functions (\ref{BintroduceF}) and (\ref{PhiExpandNearTc}) into
(\ref{BHPhiz}), we keep terms up to $0(b)$ \cite{BGRLPRD2013} to get
the equation of motion for $\chi(z)$
\begin{eqnarray}\label{BHChizEoM}
(U\chi^\prime)^\prime=\frac{2\lambda z^{1+2\Delta-d}F^{2}\zeta}{f},
\end{eqnarray}
where we have introduced a new function
\begin{eqnarray}
U(z)=\frac{e^{3b\lambda^{2}z^{4}\zeta'^{2}/2}}{z^{d-3}}.
\end{eqnarray}
Making integration of both sides of Eq. (\ref{BHChizEoM}), we have
\begin{eqnarray}\label{Chi0}
\left[\frac{\chi'(z)}{z^{d-3}}\right]\bigg|_{z\rightarrow 0}=
\left\{
\begin{array}{rl}
-\lambda\alpha_{1}\ ,~~~~~~ &  \quad {\rm for}\
(d-2)^{2}b_{n}(\lambda^{2}|_{b_{n-1}})<1, \\ \\
-\lambda(\alpha_{2A}+\alpha_{2B})  \ , &  \quad {\rm for}\
(d-2)^{2}b_{n}(\lambda^{2}|_{b_{n-1}})>1,
\end{array}\right.
\end{eqnarray}
with
\begin{eqnarray}
\alpha_{1}=\int_{0}^{1}\frac{2z^{1+2\Delta-d}F^{2}\zeta_{1}}{f}dz,~~
\alpha_{2A}=\int_{0}^{\Lambda}\frac{2z^{1+2\Delta-d}F^{2}\zeta_{2A}}{f}dz,~~
\alpha_{2B}=\int_{\Lambda}^{1}\frac{2z^{1+2\Delta-d}F^{2}\zeta_{2B}}{f}dz.
\end{eqnarray}

For clarity, we will fix the spacetime dimension $d$ in the
following discussion. Considering the case of $d=3$ and the
asymptotic behavior (\ref{infinity}), for example, near
$z\rightarrow0$ we can arrive at
\begin{eqnarray}\label{BHPhiExpandD3}
\frac{\rho}{r^{2}_+}(1-z)=\lambda\zeta(z)+\frac{\langle
\mathcal{O}\rangle^2}{r_+^{2\Delta}}\left[\chi(0)+\chi^\prime(0)z+\cdot\cdot\cdot\right].
\end{eqnarray}
From the coefficients of the $z^1$ terms in both sides of the above
formula, we can obtain
\begin{eqnarray}\label{BHRhoExpD3}
\frac{\rho}{r^{2}_{+}}=\lambda-\frac{\langle
\mathcal{O}\rangle^2}{r_+^{2\Delta}}\chi'(0),
\end{eqnarray}
where $\chi^\prime(0)$ can be easily calculated by using Eq.
(\ref{Chi0}). Therefore we will know that
\begin{eqnarray}\label{D3OExp}
\langle\mathcal{O}\rangle=\beta
T_{c}^{\Delta}\left(1-\frac{T}{T_c}\right)^{\frac{1}{2}},
\end{eqnarray}
where the coefficient $\beta$ is given by
\begin{eqnarray}\label{D3Beta}
\beta=\left\{
\begin{array}{rl}
\left(\frac{4\pi}{3}\right)^{\Delta}\sqrt{\frac{2}{\alpha_{1}}}\
,~~~~ & \quad {\rm for}\ b_{n}(\lambda^{2}|_{b_{n-1}})<1,
\\ \\
\left(\frac{4\pi}{3}\right)^{\Delta}\sqrt{\frac{2}{\alpha_{2A}+\alpha_{2B}}}\
, &  \quad {\rm for}\ b_{n}(\lambda^{2}|_{b_{n-1}})>1.
\end{array}\right.
\end{eqnarray}
Obviously, the expression (\ref{D3OExp}) is valid for different
values of the Born-Infeld parameter and scalar field mass in the
case of the ($2+1$)-dimensional superconductor. For concreteness, we
will focus on the case for the mass of the scalar field
$m^{2}L^2=-2$ and the step size $\Delta b=0.1$. Since in Ref.
\cite{JS2010} the scalar operator is given by
$\langle\mathcal{O}_{+}\rangle=\sqrt{2}\psi_{+}$ which is different
from $\langle\mathcal{O}_{+}\rangle=\psi_{+}$ in this work, we
present the condensation value $\gamma=\sqrt{2}\beta$ obtained by
the analytical S-L method and from numerical calculation with the
chosen values of the Born-Infeld parameter $b$ for the
($2+1$)-dimensional superconductor in Table \ref{D3BetaValue}. We
see that the condensation value $\gamma$ increases as the
Born-Infeld parameter $b$ increases for the fixed scalar field mass
and spacetime dimension, which indicates the consistent picture
shown in $T_{c}$ that the higher Born-Infeld electrodynamics
corrections make the condensation to be formed harder. On the other
hand, comparing with the analytical results shown in Table II of
Ref. \cite{BGRLPRD2013}, we find that the iterative procedure indeed
reduces the disparity between the analytical and numerical results.

\begin{table}[ht]
\begin{center}
\caption{\label{D3BetaValue} The condensation value
$\gamma=\sqrt{2}\beta$ obtained by the analytical S-L method and
from numerical calculation \cite{JS2010} with the chosen values of
the Born-Infeld parameter $b$ in the case of 4-dimensional AdS black
hole background. Here we fix the mass of the scalar field by
$m^{2}L^2=-2$ and the step size by $\Delta b=0.1$.}
\begin{tabular}{c c c c c c c}
         \hline \hline
$b$ & 0 & 0.1 & 0.2 & 0.3
        \\
        \hline
~~~~$Analytical$~~~~~~~~~&~~~~~~~~~$92.80$~~~~~~~~~&~~~~~~~~~$117.92$
~~~~~~~~~&~~~~~~~~~$137.22$~~~~~~~~~&~~~~~~~~~$161.14$~~~~
          \\
~~~~$Numerical$~~~~~~~~~&~~~~~~~~~$139.24$~~~~~~~~~&~~~~~~~~~$207.36$
~~~~~~~~~&~~~~~~~~~$302.76$~~~~~~~~~&~~~~~~~~~$432.64$~~~~
          \\
        \hline \hline
\end{tabular}
\end{center}
\end{table}

As another example, let us move on to the case of $d=4$. From the
asymptotic behavior (\ref{infinity}), we can expand $\phi$ when
$z\rightarrow0$ as
\begin{eqnarray}\label{BHPhiExpandD4}
\frac{\rho}{r^{3}_+}(1-z^{2})=\lambda\zeta(z)+\frac{\langle
\mathcal{O}\rangle^2}{r_+^{2\Delta}}\left[\chi(0)+\chi^\prime(0)z
+\frac{1}{2}\chi^{\prime\prime}(0)z^2+\cdot\cdot\cdot\right].
\end{eqnarray}
Considering the coefficients of $z^1$ terms in above equation, we
observe that $\chi^\prime(0)\rightarrow 0$ if $z\rightarrow 0$,
which is consistent with Eq. (\ref{Chi0}). Comparing the
coefficients of the $z^2$ terms, we have
\begin{eqnarray}\label{BHRhoExpD4}
\frac{\rho}{r^{3}_{+}}=\lambda-\frac{\langle
\mathcal{O}\rangle^2}{2r_+^{2\Delta}}\chi''(0),
\end{eqnarray}
where $\chi''(0)$ can be computed by using Eq. (\ref{Chi0}). So we
can deduce the same relation (\ref{D3OExp}) for the
($3+1$)-dimensional superconductor with the different condensation
coefficient
\begin{eqnarray}\label{D4Beta}
\beta=\left\{
\begin{array}{rl}
\pi^{\Delta}\sqrt{\frac{6}{\alpha_{1}}}\ ,~~~~ & \quad {\rm for}\
4b_{n}(\lambda^{2}|_{b_{n-1}})<1,
\\ \\
\pi^{\Delta}\sqrt{\frac{6}{\alpha_{2A}+\alpha_{2B}}}\ , & \quad {\rm
for}\ 4b_{n}(\lambda^{2}|_{b_{n-1}})>1.
\end{array}\right.
\end{eqnarray}
In Table \ref{D4BetaValue}, we give the condensation value $\beta$
obtained by the analytical S-L method with the chosen values of the
Born-Infeld parameter $b$ and step size $\Delta b$ for the
($3+1$)-dimensional superconductor. In both cases we find again
that, for the fixed scalar field mass and spacetime dimension, the
condensation value $\beta$ increases as the Born-Infeld parameter
$b$ increases, just as the observation obtained in the
($2+1$)-dimensional superconductor with Born-Infeld electrodynamics.

\begin{table}[ht]
\begin{center}
\caption{\label{D4BetaValue} The condensation value $\beta$ obtained
by the analytical S-L method with the chosen values of the
Born-Infeld parameter $b$ and step size $\Delta b$ in the case of
5-dimensional AdS black hole background. Here we fix the mass of the
scalar field by $m^{2}L^2=-3$. }
\begin{tabular}{c c c c c c c}
         \hline \hline
$b$ & 0 & 0.1 & 0.2 & 0.3
        \\
        \hline
~~~~$\Delta
b=0.05$~~~~~~~~~&~~~~~~~~~$238.91$~~~~~~~~~&~~~~~~~~~$418.95$
~~~~~~~~~&~~~~~~~~~$697.64$~~~~~~~~~&~~~~~~~~~$1195.56$~~~~
          \\
~~~~$\Delta
b=0.025$~~~~~~~~~&~~~~~~~~~$238.91$~~~~~~~~~&~~~~~~~~~$496.06$
~~~~~~~~~&~~~~~~~~~$1005.38$~~~~~~~~~&~~~~~~~~~$2303.28$~~~~
          \\
        \hline \hline
\end{tabular}
\end{center}
\end{table}

It should be noted that one can easily extend our discussion to the
higher-dimensional superconductor and get our expression
(\ref{D3OExp}), although the coefficient $\beta$ is different. Thus,
near the critical point, the scalar operator
$\langle\mathcal{O}\rangle$ will satisfy
\begin{eqnarray}\label{DOExp}
\langle\mathcal{O}\rangle\sim\left(1-T/T_c\right)^{1/2},
\end{eqnarray}
which holds for various values of the Born-Infeld parameter $b$,
spacetime dimension $d$ and mass of the scalar field $m$. It shows
that the phase transition is of the second order and the critical
exponent of the system always takes the mean-field value $1/2$. The
Born-Infeld electrodynamics will not influence the result.

\section{Conclusions}

We have generalized the variational method for the S-L eigenvalue
problem to analytically investigate the condensation and critical
phenomena of the $d$-dimensional superconductors with Born-Infeld
electrodynamics, which may help to understand the influences of the
$1/N$ or $1/\lambda$ corrections on the holographic superconductor
models. We found that the S-L method is still powerful to disclose
the properties of the holographic superconductor with various
condensates even when we take the Born-Infeld electrodynamics into
account. Using the iterative procedure in the perturbative approach
proposed by Banerjee \emph{et al.} \cite{BGRLPRD2013}, we further
improved the analytical results and the consistency with the
numerical findings for the ($2+1$)-dimensional superconductor.
Furthermore, extending the investigation to the higher-dimensional
superconductor with Born-Infeld electrodynamics, we observed again
that the analytical results derived from this method with a
reasonable step size are in very good agreement with those obtained
from numerical calculation. Our analytical result shows that the
Born-Infeld parameter makes the critical temperature of the
superconductor decrease, which can be used to back up the numerical
findings as shown in the existing literatures that the higher
Born-Infeld electrodynamics corrections can hinder the condensation
to be formed. Moreover, with the help of this analytical method, we
interestingly noted that the Born-Infeld electrodynamics, spacetime
dimension and scalar mass cannot modify the critical phenomena, and
found that the holographic superconductor phase transition belongs
to the second order and the critical exponent of the system always
takes the mean-field value. It should be noted that one can easily
extend our technique to the holographic superconductor models with
the logarithmic form \cite{JPCPLB} and exponential form
\cite{ZPCJNPB} of nonlinear electrodynamics. More recently, a model
of p-wave holographic superconductors from charged Born-Infeld black
holes \cite{CSJHEP2015} via a Maxwell complex vector field model
\cite{CaiPWave-1,CaiPWave-2,CaiPWave-3} was studied numerically. It
would be of interest to generalize our study to this p-wave model
and analytically discuss the effect of the Born-Infeld
electrodynamics on the system. We will leave it for further study.

\begin{acknowledgments}

This work was supported by the National Natural Science Foundation
of China under Grant Nos. 11275066, 11175065 and 11475061; Hunan
Provincial Natural Science Foundation of China under Grant Nos.
12JJ4007 and 11JJ7001; and FAPESP No. 2013/26173-9.

\end{acknowledgments}

\end{document}